\documentclass[10pt,twocolumn]{article}

\usepackage[utf8]{inputenc}
\usepackage[T1]{fontenc}
\usepackage[margin=2cm]{geometry}
\usepackage{booktabs}
\usepackage{subcaption}
\usepackage{listings}
\usepackage{xcolor}
\usepackage{enumitem}
\usepackage{multirow}
\usepackage{tabularx}
\usepackage{tikz}
\usetikzlibrary{positioning,arrows.meta,fit,backgrounds,calc}
\usepackage{orcidlink}
\usepackage{hyperref}
\usepackage[numbers,sort&compress]{natbib}
\usepackage{authblk}

\hypersetup{
  pdftitle={Codebase-Memory: Tree-Sitter-Based Knowledge Graphs for LLM Code Exploration via MCP},
  pdfauthor={Martin Vogel, Falk Meyer-Eschenbach, Severin Kohler, Elias Grünewald, Felix Balzer},
  pdfsubject={Software Engineering, LLM Agents, Knowledge Graphs},
  pdfkeywords={knowledge graph, LLM agents, code exploration, Tree-Sitter, Model Context Protocol, MCP, static analysis},
}

\usepackage{titlesec}
\titlespacing*{\section}{0pt}{1.5ex plus 0.5ex minus 0.2ex}{0.8ex plus 0.2ex}
\titlespacing*{\subsection}{0pt}{1.2ex plus 0.4ex minus 0.2ex}{0.5ex plus 0.1ex}

\usepackage{etoolbox}
\makeatletter
\patchcmd{\@maketitle}{\vskip 2em}{\vskip 0.5em}{}{}
\makeatother

\title{Codebase-Memory: Tree-Sitter-Based Knowledge Graphs\\ for LLM Code Exploration via MCP}

\author[1]{Martin Vogel\,\orcidlink{0009-0003-5148-2900}}
\author[2,3,4]{Falk Meyer-Eschenbach\,\orcidlink{0009-0000-9813-0249}}
\author[5,6]{Severin Kohler\,\orcidlink{0000-0002-7718-6187}}
\author[2]{\\Elias Grünewald\,\orcidlink{0000-0001-9076-9240}}
\author[2]{Felix Balzer\,\orcidlink{0000-0003-1575-2056}}

\affil[1]{Independent Researcher, Berlin, Germany}
\affil[2]{Institute of Medical Informatics, Charit\'{e} -- Universit\"{a}tsmedizin Berlin, Berlin, Germany}
\affil[3]{Clinical Study Center, Berlin Institute of Health, Berlin, Germany}
\affil[4]{Institute of Informatics, Humboldt University, Berlin, Germany}
\affil[5]{Institute of Informatics, Freie Universit\"{a}t Berlin, Berlin, Germany}
\affil[6]{Health Data Science Unit, University Hospital Heidelberg, Heidelberg, Germany}

\date{}

\begin{document}
\maketitle


\begin{abstract}
\noindent
Large Language Model (LLM) coding agents typically explore codebases through repeated file-reading and grep-searching, consuming thousands of tokens per query without structural understanding. We present \textsc{Codebase-Memory}, an open-source system that constructs a persistent, Tree-Sitter-based knowledge graph via the Model Context Protocol~(MCP), parsing 66 languages through a multi-phase pipeline with parallel worker pools, call-graph traversal, impact analysis, and community discovery \cite{blondel2008louvain}. Evaluated across 31 real-world repositories, \textsc{Codebase-Memory} achieves 83\% answer quality versus 92\% for a file-exploration agent, at ten times fewer tokens and 2.1 times fewer tool calls. For graph-native queries such as hub detection and caller ranking, it matches or exceeds the explorer on 19 of 31 languages.
\end{abstract}

\section{Introduction}
\label{sec:introduction}

The emergence of LLM-based coding agents, such as Claude Code~\citep{anthropic2024mcp}, Cursor, and Aider~\citep{gauthier2023aider}, has transformed software development by enabling natural-language-driven code exploration, bug fixing, and refactoring. These agents interact with codebases through tool calls, reading files, searching for patterns, and listing directory contents. While effective for small-scale tasks, this text-based exploration strategy scales poorly. A typical codebase exploration session requires dozens of tool calls, consuming hundreds of thousands of tokens before the agent develops sufficient understanding to answer a structural question such as ``What breaks if I change this function?''~\citep{qiu2025locobench}.

This inefficiency stems from a fundamental mismatch: LLM agents operate on unstructured text, yet the questions developers ask are inherently structural---call graphs, dependency chains, module boundaries, and impact analysis. Text-based search cannot capture transitive relationships without iteratively following references, each iteration consuming additional tokens and increasing the risk of lost context~\citep{liu2024lostmiddle}. Recent empirical analyses confirm that input tokens dominate the cost of agentic coding tasks, even with caching~\citep{hrubec2025tokenusage}.

Recent work on coding agents~\citep{yang2024sweagent,zhang2024autocoderover,xia2024agentless} has focused on improving agent strategies, better prompting, hierarchical localization, or fault-localization heuristics, while leaving the underlying code retrieval mechanism largely unchanged. Concurrently, graph-based code representations such as Code Property Graphs~\citep{yamaguchi2014cpg} and CodeQL~\citep{avgustinov2016ql} have proven powerful for static analysis but remain heavyweight, requiring specialized databases and query languages not optimized for LLM consumption. Emerging work on repository-aware knowledge graphs~\citep{yang2025kgcompass} and graph-enhanced retrieval agents~\citep{shah2025ranger} demonstrates the growing interest in structural code representations for LLM agents, though these systems typically require complex infrastructure. The cost of this inefficiency is concrete: production agentic workloads report per-task LLM API costs of several dollars on non-trivial repositories~\citep{hrubec2025tokenusage,yang2024sweagent}. The recent emergence of the Model Context Protocol (MCP)~\citep{anthropic2024mcp} as an open standard for connecting LLM agents to external tools presents an opportunity to expose structural code queries as lightweight, agent-native tools---yet no existing solution exploits this to combine structural query capability with zero-infrastructure deployment.

To bridge this gap, we propose treating codebase structure as a first-class, queryable knowledge graph exposed directly to LLM agents via lightweight structural tools rather than raw file content. We instantiate this in our implementation \textsc{Codebase-Memory}, which parses codebases using Tree-Sitter~\citep{brunsfeld2018treesitter}, stores the resulting graph in SQLite across 66 languages, and exposes 14 structural query tools via MCP~\citep{anthropic2024mcp}, maintained incrementally through file-watching and content-hash-based re-indexing. The system ships as a single statically linked C binary with zero runtime dependencies.

Our contributions are:
\begin{itemize}[nosep]
  \item A \textbf{knowledge-graph architecture} for code that combines Tree-Sitter parsing across 66 languages, a multi-phase build pipeline with parallel extraction, 6-strategy call resolution, and Louvain community detection, stored in a single SQLite file with zero external dependencies.
  \item An \textbf{MCP-based tool interface} exposing 14 typed structural queries (call-path tracing, impact analysis, hub detection) to any MCP-compatible LLM agent, with sub-millisecond query latency.
  \item A \textbf{head-to-head evaluation} across 31 languages showing ten times lower token cost and 2.1 times fewer tool calls at competitive quality (83\% vs.\ 92\%), with a systematic analysis of where graph-based retrieval excels and where file-based exploration remains necessary.
\end{itemize}

The remainder of this paper is structured as follows: Section~\ref{sec:background} surveys related work in structural code analysis, code retrieval for LLMs, and token efficiency. Section~\ref{sec:system} describes the system architecture, graph schema, pipeline, and security hardening. Section~\ref{sec:evaluation} presents the benchmark evaluation, performance measurements, and adoption metrics. Section~\ref{sec:discussion} discusses trade-offs, threats to validity, and future work. Section~\ref{sec:conclusion} concludes.
\section{Related Work}
\label{sec:background}

\subsection{Structural Code Analysis}
\label{sec:bg-structural}

Structural code representations have a long history in software engineering. Program Dependence Graphs~\citep{ferrante1987pdg} unify data and control dependencies. Code Property Graphs~\citep{yamaguchi2014cpg} merge Abstract Syntax Trees (ASTs), control flow graphs, and PDGs into a single queryable structure. CodeQL~\citep{avgustinov2016ql} provides a declarative query language over relational code databases. While powerful, these systems are heavyweight, requiring specialized databases and domain-specific query languages, and are not designed for LLM consumption.

Tree-Sitter~\citep{brunsfeld2018treesitter} provides fast, incremental, error-tolerant parsing with grammars for more than 100 languages. It has become the de facto standard for editor-integrated code analysis and has been adopted by tools like Aider~\citep{gauthier2023aider} for repository mapping. Recent work on AST-based chunking~\citep{zhang2025cast} demonstrates that preserving syntactic structure during code retrieval improves both recall and downstream task performance compared to naive text chunking.

\subsection{Code Retrieval for LLMs}
\label{sec:bg-rag}

Repository-level code understanding remains challenging for LLMs. RepoCoder~\citep{zhang2023repocoder} employs iterative BM25 retrieval for code completion. DocPrompting~\citep{zhou2023docprompting} retrieves documentation as generation context. A recent survey~\citep{racgsurvey2025} identifies structural retrieval as a key frontier.

Several recent systems construct code graphs for LLM-based retrieval. GraphCoder~\citep{liu2024graphcoder} built Code Context Graphs for repository-level code completion (ASE 2024). CodexGraph~\citep{liu2024codexgraph} exposed code graphs to LLM agents via graph database interfaces (NAACL 2025). KGCompass~\citep{yang2025kgcompass} linked issues and code entities in repository-aware knowledge graphs, achieving 58.3\% on SWE-bench Lite. RepoGraph~\citep{ouyang2025repograph} constructed repository-level code graphs that boost existing agents by 32.8\% relative improvement on SWE-bench (ICLR 2025).

A second line of work focuses on graph-guided agent navigation. LocAgent~\citep{chen2025locagent} parsed codebases into directed heterogeneous graphs for multi-hop code localization (ACL 2025). GraphCodeAgent~\citep{li2025graphcodeagent} constructed dual requirement-structural graphs for retrieval-augmented code generation. RANGER~\citep{shah2025ranger} used graph-enhanced retrieval for repository-level queries. Prometheus~\citep{pan2025prometheus} combined Tree-Sitter-based knowledge graphs with unified memory for multilingual issue resolution. The Repository Intelligence Graph~\citep{cherny2026rig} provided a deterministic architectural map for LLM code assistants.

More fundamentally, SemanticForge~\citep{zhang2025semanticforge} employed dual static-dynamic knowledge graphs with neural graph query generation, and the Code Graph Model~\citep{cgm2025codegraphmodel} integrated graph representations directly into LLM attention mechanisms.

Our work differs from prior approaches by using MCP as a standardized interface compatible with any MCP-capable agent, SQLite for zero-dependency deployment, incremental synchronization for live workflows, and support for 66 languages via a single binary.

\subsection{LLM Coding Agents}
\label{sec:bg-agents}

SWE-bench~\citep{jimenez2024swebench} established the gold-standard benchmark for coding agents. SWE-Agent~\citep{yang2024sweagent} introduced Agent-Computer Interfaces (ACIs) optimized for LLM interaction. AutoCodeRover~\citep{zhang2024autocoderover} combines AST-aware code search with fault localization. Agentless~\citep{xia2024agentless} demonstrated that a simple three-phase approach achieves competitive results without an agentic loop. OpenHands~\citep{wang2024openhands} provides an extensible open-source platform.

These systems optimize the agent strategy while using simple text-based retrieval. Our work is orthogonal: we optimize the retrieval layer, and our approach can be combined with any of these agent architectures.

\subsection{Token Efficiency}
\label{sec:bg-tokens}

LLMLingua~\citep{jiang2023llmlingua} achieves up to 20 times prompt compression via perplexity-based token pruning (EMNLP 2023), extended by LLMLingua-2~\citep{pan2024llmlingua2} with task-agnostic data distillation (Findings of ACL 2024). Liu et al.~\citep{liu2024lostmiddle} show that LLMs underutilize information in the middle of long contexts. LoCoBench-Agent~\citep{qiu2025locobench} reveals a comprehension-efficiency tradeoff: thorough codebase exploration conflicts with token efficiency, and agents cluster on a Pareto frontier. Our approach addresses this tradeoff by providing structurally relevant information from the start, avoiding unnecessary token consumption while maintaining comprehension quality.

\section{System Design}
\label{sec:system}

\subsection{Architecture Overview}
\label{sec:sys-arch}

The architecture follows a three-stage pipeline. First, the \textbf{Parse} stage walks Tree-Sitter ASTs across 66 languages to extract definitions (functions, methods, classes, interfaces, enums, and types with signatures, return types, receivers, decorators, complexity, and export status), call sites, imports (8 language-specific parsers plus a generic fallback), references, and trait implementations. For Go, C, and C++, a hybrid approach augments Tree-Sitter extraction with LSP-style type resolution to improve call-graph accuracy in the presence of method receivers, pointer indirection, and package-qualified identifiers. Second, the \textbf{Build} stage executes a multi-phase pipeline (Section~\ref{sec:sys-pipeline}) with parallel worker pools that write extracted entities to per-worker in-memory graph buffers, merge them, and flush to SQLite with deferred index creation. Third, the \textbf{Serve} stage exposes the graph via an MCP server with 14 typed tools that LLM agents invoke through standard tool-call semantics.

\textsc{Codebase-Memory} is implemented as a single, statically linked C binary with zero runtime dependencies. The system vendors 66 Tree-Sitter grammars as C source and compiles to a self-contained executable for macOS, Linux, and Windows. Pipeline orchestration, graph storage, MCP protocol, Cypher query engine, and background synchronization are implemented entirely in C. All state lives in a single SQLite file.

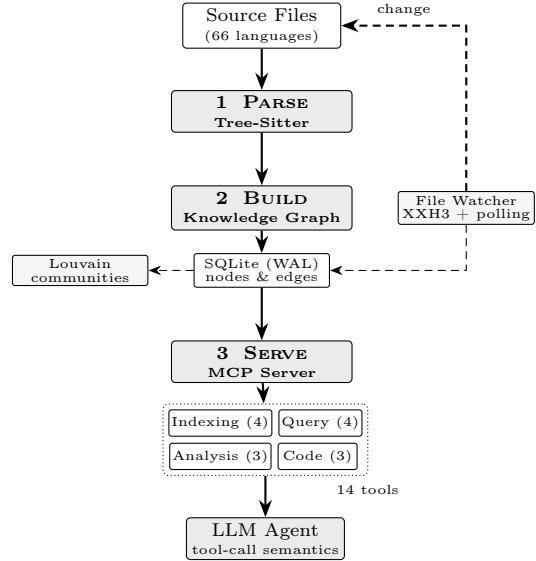
\begin{figure}[t]
\centering
\begin{tikzpicture}[
  node distance=0.55cm and 0.3cm,
  >=Stealth,
  every node/.style={font=\scriptsize},
  box/.style={
    draw, rectangle, rounded corners=1.5pt,
    minimum width=2.1cm, minimum height=0.55cm,
    align=center, inner sep=2pt
  },
  stagebox/.style={
    box, fill=black!8, minimum width=2.4cm, font=\scriptsize\bfseries
  },
  smallbox/.style={
    draw, rectangle, rounded corners=1pt,
    minimum width=1.8cm, minimum height=0.4cm,
    align=center, inner sep=1.5pt, font=\tiny
  },
  catbox/.style={
    draw, rectangle, rounded corners=1pt,
    minimum height=0.35cm, minimum width=1.05cm,
    align=center, inner sep=1pt, font=\tiny
  },
  arr/.style={->, thick},
  darr/.style={->, densely dashed, thin}
]

\node[box] (src) {Source Files\\[-1pt]{\tiny(66 languages)}};

\node[stagebox, below=of src] (parse) {%
  \textsc{1\;\;Parse}\\[-1pt]{\tiny Tree-Sitter}};

\node[stagebox, below=0.7cm of parse] (build) {%
  \textsc{2\;\;Build}\\[-1pt]{\tiny Knowledge Graph}};

\node[smallbox, below=0.3cm of build] (sqlite) {%
  SQLite (WAL)\\[-1pt] nodes \& edges};

\node[stagebox, below=0.7cm of sqlite] (serve) {%
  \textsc{3\;\;Serve}\\[-1pt]{\tiny MCP Server}};

\node[catbox, below=0.35cm of serve, xshift=-0.55cm] (t1) {Indexing (4)};
\node[catbox, right=0.08cm of t1] (t2) {Query (4)};
\node[catbox, below=0.08cm of t1] (t3) {Analysis (3)};
\node[catbox, right=0.08cm of t3] (t4) {Code (3)};

\begin{scope}[on background layer]
  \node[draw, densely dotted, rounded corners=2pt, inner sep=2pt,
        fit=(t1)(t2)(t3)(t4), label={[font=\tiny,anchor=north west,xshift=0.3cm]
        below right:{\,14 tools}}] (toolfit) {};
\end{scope}

\node[box, fill=black!8, below=0.55cm of toolfit]
  (llm) {LLM Agent\\[-1pt]{\tiny tool-call semantics}};

\node[smallbox, right=0.6cm of build, fill=black!4] (watch) {%
  File Watcher\\[-1pt]{\tiny XXH3 + polling}};

\node[smallbox, left=0.6cm of sqlite, fill=black!4] (louv) {%
  Louvain\\[-1pt]{\tiny communities}};


\draw[arr] (src) -- (parse);
\draw[arr] (parse) -- (build);
\draw[arr] (build) -- (sqlite);
\draw[arr] (sqlite) -- (serve);
\draw[arr] (serve) -- (toolfit);
\draw[arr] (toolfit) -- (llm);

\draw[arr, densely dashed] (watch.north) -- (watch.north |- src)
  -- node[above, font=\tiny] {change} (src.east);
\draw[darr] (watch.south) -- (watch.south |- sqlite.east) -- (sqlite.east);

\draw[darr] (sqlite.west) -- (louv.east);

\end{tikzpicture}
\caption{\textsc{Codebase-Memory} architecture. Source files are parsed via Tree-Sitter, stored as a knowledge graph in SQLite, and exposed to LLM agents through 14 typed MCP tools. A file watcher triggers incremental re-indexing on changes.}
\label{fig:architecture}
\end{figure}

A background file watcher monitors the repository for changes using adaptive polling. When files are modified, an XXH3 content hash triggers incremental re-indexing of only the affected files. Figure~\ref{fig:architecture} illustrates the overall architecture.

\subsection{Graph Schema}
\label{sec:sys-schema}

The knowledge graph uses a property-graph model with typed nodes and edges:

\begin{table}[ht]
\caption{Node types in the knowledge graph.}
\label{tab:nodes}
\small
\begin{tabularx}{\columnwidth}{@{}lX@{}}
\toprule
\textbf{Node Type} & \textbf{Extracted From} \\
\midrule
\texttt{Project}, \texttt{Package}, \texttt{Folder} & Directory structure \\
\texttt{File}, \texttt{Module} & File system \\
\texttt{Function}, \texttt{Method}, \texttt{Class} & Tree-Sitter AST \\
\texttt{Interface}, \texttt{Enum}, \texttt{Type} & Tree-Sitter AST \\
\texttt{Route} & Framework detection \\
\bottomrule
\end{tabularx}
\end{table}

\begin{table}[ht]
\caption{Edge types representing code relationships.}
\label{tab:edges}
\footnotesize
\begin{tabularx}{\columnwidth}{@{}Xl@{}}
\toprule
\textbf{Edge Type} & \textbf{Semantics} \\
\midrule
\texttt{CALLS}, \texttt{HTTP\_CALLS}, \texttt{ASYNC\_CALLS} & Invocation \\
\texttt{IMPORTS} & Module import \\
\texttt{CONTAINS\_*}, \texttt{DEFINES}, \texttt{DEFINES\_METHOD} & Structural nesting \\
\texttt{IMPLEMENTS} & Interface impl. \\
\texttt{HANDLES} & Route$\to$handler \\
\texttt{INHERITS}, \texttt{DECORATES} & OOP hierarchy \\
\texttt{USES\_TYPE}, \texttt{USAGE} & Symbol reference \\
\texttt{THROWS}, \texttt{READS}, \texttt{WRITES} & Side effects \\
\texttt{CONFIGURES} & Config linkage \\
\texttt{TESTS}, \texttt{FILE\_CHANGES\_WITH} & Semantic links \\
\texttt{MEMBER\_OF} & Community membership \\
\bottomrule
\end{tabularx}
\end{table}

The schema includes \texttt{HTTP\_CALLS} and \texttt{ASYNC\_CALLS} edges discovered via cross-service HTTP route/call-site matching (6 framework-specific extractors covering Python, Go, Java/Spring, Kotlin/Ktor, Express.js, and Laravel) with confidence scoring (0.0--1.0), enabling REST endpoints to be treated as first-class graph entities. This allows the system to represent distributed codebases, such as microservice architectures, across multiple languages.

\subsection{Multi-Pass Pipeline}
\label{sec:sys-pipeline}

The pipeline executes in six phases within a single SQLite transaction. Phases~1--4 write to an in-memory graph buffer (\texttt{cbm\_gbuf\_t}), a C struct holding nodes and edges in hash maps indexed by qualified name, label, and ID, avoiding SQLite overhead during bulk insertion. Extraction and resolution phases dispatch work to a pthreads-based worker pool with atomic work-stealing: each worker writes to a per-worker buffer, which are merged after completion. The buffer assigns temporary sequential IDs that are remapped to real SQLite row IDs during flush. Phase~6 operates directly on the database after index creation.

\begin{table}[ht]
\caption{Pipeline phases and their outputs.}
\label{tab:pipeline}
\footnotesize
\begin{tabularx}{\columnwidth}{@{}lX@{}}
\toprule
\textbf{Phase} & \textbf{Output} \\
\midrule
1.\;Structure    & File discovery; Project, Package, Folder, File nodes + containment edges \\
2.\;Extraction   & Parallel definition extraction via pthreads worker pool; Function, Method, Class, Interface, Enum, Type nodes; decorator tags; FunctionRegistry \\
3.\;Resolution   & Parallel call, usage, and semantic resolution; CALLS, IMPORTS, USAGES, USES\_TYPE, IMPLEMENTS, INHERITS, DECORATES edges \\
4.\;Enrichment   & TESTS edges, HTTP route matching, config linking, git co-change edges \\
5.\;Flush        & Bulk INSERT into SQLite with deferred index creation \\
6.\;Post-index   & Louvain communities, XXH3 file hashes \\
\bottomrule
\end{tabularx}
\end{table}

\subsection{Call Resolution}
\label{sec:sys-callres}

Resolving raw callee names (e.g., \texttt{pkg.Func}) to qualified graph nodes is the core linking challenge. The \texttt{FunctionRegistry} indexes all definitions by qualified name (exact map) and simple name (reverse index). Resolution uses a prioritized 6-strategy cascade with per-strategy confidence scoring:

\begin{enumerate}[nosep]
  \item \textbf{Import map} (confidence 0.95): Split callee into prefix.suffix; look up prefix in the file's import map to obtain the module qualified name; join with suffix; exact match in registry.
  \item \textbf{Import map suffix} (0.85): Fallback when the exact import-map match fails; attempt suffix-based matching against import-resolved module paths.
  \item \textbf{Same module} (0.90): Prefix callee with the enclosing file's module qualified name; exact match.
  \item \textbf{Unique name} (0.75): Look up simple name in the reverse index; accept if exactly one candidate project-wide (penalized if not import-reachable).
  \item \textbf{Suffix match} (0.55): Among multiple candidates, select by suffix match with import-distance scoring; nearest module path wins.
  \item \textbf{Fuzzy} (0.30--0.40): Last-resort matching using string similarity when no structured resolution succeeds.
\end{enumerate}

From our observations, strategies~1--3 resolve {$\sim$}80\% of calls in well-structured codebases, while strategies~4--6 handle cross-module references and dynamic dispatch.

\paragraph{LSP-Style Hybrid Type Resolution.}
The name-based cascade above operates on string-level callee tokens and does not track expression types. This limits accuracy for languages with method receivers (Go), pointer indirection and implicit \texttt{this} (C/C++), or template-dependent calls (C++). To address this, the system includes dedicated type-resolution passes for Go, C, and C++ that run after the initial Tree-Sitter extraction phase.

Each pass builds a per-file \emph{TypeRegistry} populated with all definitions extracted in the Build stage (both file-local and cross-file), augmented with auto-generated standard-library type stubs. A \emph{Scope} structure tracks variable bindings introduced by declarations, assignments, short-variable declarations (Go), and function parameters. The resolver then walks all call-site AST nodes and evaluates the receiver expression's type bottom-up: identifier lookup in the scope chain, field and method lookup with base-class or embedded-type traversal, return-type propagation through call chains, and type simplification (reference unwrapping, pointer dereferencing, alias resolution). For C++, the pass additionally handles namespace resolution (\texttt{using} directives, declarations, and aliases), template parameter defaults, and pending template calls that are resolved retroactively when concrete argument types become available at call sites. Resolved calls carry the fully qualified callee name and bypass the string-based cascade entirely, producing higher-confidence edges in the graph. The passes operate in batch mode across all files of the respective language, enabling cross-file resolution within a single pipeline run.

\subsection{MCP Tool Interface}
\label{sec:sys-tools}

The system exposes 14 tools grouped into four categories:

\begin{table}[ht]
\caption{MCP tools exposed by \textsc{Codebase-Memory}.}
\label{tab:tools}
\small
\begin{tabularx}{\columnwidth}{@{}llX@{}}
\toprule
\textbf{Category} & \textbf{Tool} & \textbf{Description} \\
\midrule
\multirow{4}{*}{Indexing} & \texttt{index\_repository} & Build/update graph \\
& \texttt{index\_status} & Poll indexing progress \\
& \texttt{list\_projects} & List indexed repos \\
& \texttt{delete\_project} & Remove index \\
\midrule
\multirow{4}{*}{Query} & \texttt{search\_graph} & Symbol search \\
& \texttt{trace\_call\_path} & Call-chain traversal \\
& \texttt{query\_graph} & Cypher-like queries \\
& \texttt{ingest\_traces} & Import runtime traces \\
\midrule
\multirow{3}{*}{Analysis} & \texttt{detect\_changes} & Git diff impact \\
& \texttt{get\_graph\_schema} & Schema introspection \\
& \texttt{get\_architecture} & Architecture summary \\
\midrule
\multirow{3}{*}{Code} & \texttt{get\_code\_snippet} & Source retrieval \\
& \texttt{search\_code} & Full-text search \\
& \texttt{manage\_adr} & Architecture decisions \\
\bottomrule
\end{tabularx}
\end{table}

Each tool returns structured JSON that the LLM agent processes directly. The \texttt{query\_graph} tool supports a Cypher-like query language for arbitrary graph traversals, while \texttt{trace\_call\_path} provides directional (inbound/outbound) call-chain tracing with configurable depth.

\subsection{Incremental Synchronization}
\label{sec:sys-sync}

On each file-system event, the system computes an XXH3 hash of the modified file and compares it against the stored hash. If changed, the file's nodes and edges are deleted and re-parsed with Tree-Sitter; the hash is updated and affected Louvain community assignments are re-computed. XXH3 is a non-cryptographic 64-bit hash achieving {$\sim$}30\,GB/s throughput, chosen over cryptographic hashes for its speed in content-addressed indexing where collision resistance is not a security requirement.

\subsection{Community Detection}
\label{sec:sys-community}

The system applies Louvain modularity optimization~\citep{blondel2008louvain} to partition the call graph into functional communities. The algorithm iterates two phases: (a)~\emph{local moving}, each node greedily joins the neighboring community maximizing modularity gain $\Delta Q = w_{\mathrm{in}} - \gamma \cdot k_i \cdot \Sigma_{\mathrm{tot}} / (2m)$, where $\gamma{=}1.0$ is the resolution parameter, $k_i$ is the node's weighted degree, $\Sigma_{\mathrm{tot}}$ is the community's total degree, and $m$ is total edge weight; (b)~\emph{refinement}, communities with internal density ${<}1\%$ are split by ejecting weakly-connected members. Convergence typically occurs in 3--5 iterations. The algorithm operates on \texttt{CALLS}, \texttt{HTTP\_CALLS}, and \texttt{ASYNC\_CALLS} edges, producing Community nodes and \texttt{MEMBER\_OF} edges used by \texttt{get\_architecture}.

\subsection{Security Hardening}
\label{sec:sys-security}

MCP servers present a distinctive trust challenge: they execute with the host agent's full permissions, yet users install them as opaque binaries from third-party repositories. As LLM agents increasingly operate with broad file-system and network access, the integrity of every tool in the agent's toolchain becomes a supply-chain security concern. A compromised or malicious MCP server could exfiltrate source code, inject backdoors, or tamper with developer environments, all while appearing to function normally. This threat model motivates the defense-in-depth approach described below.

\paragraph{8-Layer CI Audit Suite.}
An automated audit suite runs on every commit before merge:
\begin{enumerate}[nosep]
  \item \emph{Static allow-list audit:} Every call to dangerous libc functions (\texttt{system}, \texttt{popen}, \texttt{fork}, \texttt{execvp}) must appear in an audited allow-list with explicit justification; new calls fail CI.
  \item \emph{Binary string audit:} Post-build scanning of the compiled binary for hardcoded URLs (only GitHub API and localhost permitted), embedded credentials, and suspicious base64-encoded payloads.
  \item \emph{Network egress monitoring:} On Linux, the MCP session runs under \texttt{strace} to capture all \texttt{connect()} syscalls; only localhost, DNS, and the GitHub release API are permitted.
  \item \emph{Install output path validation:} A sandboxed dry-run verifies that the installer writes only to expected directories and blocks writes to sensitive paths (\texttt{\textasciitilde/.ssh}, \texttt{\textasciitilde/.gnupg}, \texttt{\textasciitilde/.aws}).
  \item \emph{Smoke-test hardening:} End-to-end functional tests verify indexing, querying, and clean shutdown with no residual processes.
  \item \emph{Graph-UI audit:} Frontend asset scanning blocks external domains, tracking scripts, and hidden iframes; the HTTP server binds to \texttt{127.0.0.1} only with locked CORS policy.
  \item \emph{MCP robustness testing:} 23 adversarial JSON-RPC payloads cover malformed JSON, SQL injection (\texttt{DROP TABLE}, \texttt{ATTACH DATABASE}), shell injection (\texttt{\$(whoami)}, pipe sequences), path traversal, ReDoS patterns, and oversized inputs; no payload may cause a crash or hang.
  \item \emph{Vendored dependency integrity:} SHA-256 checksums for all 72 vendored library files (including 66 Tree-Sitter grammars) detect supply-chain tampering; vendored code is additionally scanned for dangerous calls.
\end{enumerate}

\paragraph{Code-Level Protections.}
Shell arguments are validated via \texttt{cbm\_validate\_shell\_arg()} before all \texttt{popen} calls, rejecting metacharacters, backticks, and substitution patterns. A SQLite authorizer callback blocks \texttt{ATTACH}/\texttt{DETACH} statements at the engine level, preventing SQL-injection-based file creation. The \texttt{get\_code\_snippet} tool applies \texttt{realpath()} containment checks to prevent path-traversal reads outside the project root. All tests are compiled with AddressSanitizer and UndefinedBehaviorSanitizer; a 15-minute soak test under ASan stress-tests sustained workloads for memory safety regressions.

\paragraph{Release Verification Pipeline.}
Releases follow a three-stage draft--verify--publish model that integrates established industry security tooling. In the \emph{draft} stage, binaries are built, signed via Sigstore cosign (Linux Foundation), and attested with SLSA build provenance (Google/OpenSSF). GitHub CodeQL (Microsoft) runs static application security testing on every push, blocking the release on any open alert. In the \emph{verify} stage, every binary is submitted to VirusTotal (Google/Chronicle) via the \texttt{ghaction-virustotal} GitHub Action, where more than 70 antivirus engines scan for malware; a zero-tolerance policy blocks publication if any engine flags a detection. The CI pipeline polls each scan until all engines have completed (up to 20 minutes), requiring a minimum of 60 engines to report. Additionally, platform-native antivirus scans run during CI smoke tests on every build: Windows Defender (\texttt{MpCmdRun.exe} with ML heuristics and up-to-date signatures) on Windows, and ClamAV (Cisco/Talos, with \texttt{freshclam} signature updates) on Linux and macOS. An OpenSSF Scorecard gate (Google/OSSF) enforces a minimum repository health score. Only after all gates pass does the release become public. Each release includes SHA-256 checksums, a CycloneDX software bill of materials (SBOM) listing all seven vendored dependencies, and instructions for independent user verification via \texttt{gh attestation verify} and \texttt{cosign verify-blob}. To our knowledge, this level of automated binary verification, combining multi-engine antivirus scanning from Google, Microsoft, and Cisco with cryptographic build provenance and zero-tolerance release gating, is not commonly implemented by open-source MCP servers or developer tools distributed via GitHub.

\section{Evaluation}
\label{sec:evaluation}

We evaluate \textsc{Codebase-Memory} along four dimensions: (1)~a head-to-head comparison of an MCP-augmented agent against a file-exploration agent, (2)~a qualitative analysis of where each approach excels, (3)~a measurement of system performance, and (4)~early community adoption as a proxy for practical relevance.

\subsection{Head-to-Head Benchmark}
\label{sec:eval-benchmark}

We designed a benchmark consisting of 12 standardized question categories covering hub detection, caller ranking, dependency manifests, and full call-chain tracing (Table~\ref{tab:qcategories}). Each of 31 programming languages was tested against a real open-source repository ranging from 78 nodes (HCL/Terraform) to 49,398 nodes (Python/Django). Two agents were compared: an MCP Agent with access to \textsc{Codebase-Memory}'s 14 tools, and an Explorer Agent using conventional file-reading and grep-based exploration. Both agents used Claude Opus 4.6 as the LLM backend and received identical questions. Responses were graded by the first author against reference answers derived from manual code inspection. Scores ${\geq}0.80$ were classified as PASS, 0.40--0.79 as PARTIAL, and ${<}0.40$ as FAIL.

\begin{table}[ht]
\caption{Benchmark question categories and primary MCP tools.}
\label{tab:qcategories}
\small
\begin{tabularx}{\columnwidth}{@{}clX@{}}
\toprule
\textbf{Q\#} & \textbf{Category} & \textbf{Primary Tool} \\
\midrule
1 & Indexing & \texttt{get\_graph\_schema} \\
2--3 & Discovery & \texttt{search\_graph} \\
4 & Pattern matching & \texttt{search\_graph} \\
5 & Code retrieval & \texttt{get\_code\_snippet} \\
6 & Code search & \texttt{search\_code} \\
7--8 & Call tracing & \texttt{trace\_call\_path} \\
9--10 & Graph query & \texttt{query\_graph} \\
11 & OOP analysis & \texttt{query\_graph} \\
12 & File operations & \texttt{get\_architecture} \\
\bottomrule
\end{tabularx}
\end{table}

Table~\ref{tab:headtohead} summarizes the head-to-head comparison. The MCP Agent achieves 90\% of the Explorer Agent's quality score while consuming an order of magnitude fewer tokens and requiring 2.1 times fewer tool calls.

\begin{table}[ht]
\caption{Head-to-head comparison: MCP Agent vs.\ Explorer Agent.}
\label{tab:headtohead}
\footnotesize
\begin{tabular*}{\columnwidth}{@{\extracolsep{\fill}}lrrr@{}}
\toprule
\textbf{Metric} & \textbf{MCP} & \textbf{Explorer} & \textbf{Difference} \\
\midrule
Quality score      & 0.83    & 0.92     & 90\% of Explorer \\
Tool calls / question & 2.3  & 4.8      & 2.1 times fewer \\
Tokens / question  & {$\sim$}1,000 & {$\sim$}10,000 & 10 times fewer \\
Query latency      & $<$1\,ms & 10--30\,s & $>$100 times faster \\
\bottomrule
\end{tabular*}
\end{table}

The MCP Agent shows advantages in hub detection and caller ranking for 19 of 31 languages---queries that require following pre-materialized graph edges. The strongest results appear for functional languages (Haskell, OCaml, Elixir), where the quality gap narrows to {$\sim$}1\%.

The Explorer retains an advantage for full source context (16/31 languages) and exhaustive call-site grep (10/31)---queries requiring line-level code that the graph intentionally does not store. The weakest MCP result is macro-heavy C (0.58 vs.\ 1.00), where macros are not represented in the AST.

The speed difference arises because the MCP Agent resolves structural queries via pre-computed graph lookups (breadth-first search via SQL recursive Common Table Expression: {$\sim$}0.3\,ms). The Explorer Agent must discover structure at query time: grep for function names, read matching files, parse context, repeat. This multiplies tool calls and tokens linearly with codebase size. The graph approach pays the indexing cost once (6\,s for 49K nodes) and amortizes it across all subsequent queries.

\subsection{Comparison with Alternative Approaches}
\label{sec:eval-comparison}

Table~\ref{tab:comparison} positions \textsc{Codebase-Memory} against the three dominant paradigms for LLM code retrieval.

\begin{table}[ht]
\caption{Comparison of code retrieval approaches for LLM agents.}
\label{tab:comparison}
\scriptsize
\begin{tabularx}{\columnwidth}{@{}l*{4}{>{\centering\arraybackslash}X}@{}}
\toprule
\textbf{Feature} & \textbf{Emb./RAG} & \textbf{Repo-Map} & \textbf{Graph+LLM} & \textbf{Ours} \\
\midrule
Languages       & 10--30 & {$\sim$}100 & 8--14  & \textbf{66} \\
Struct.\ queries & No  & No   & Yes  & \textbf{Yes} \\
Infra.          & Vector DB  & None & Neo4j  & \textbf{SQLite} \\
Persistence     & Yes  & No   & Yes  & \textbf{Yes} \\
Embed.\ model   & Yes  & No   & Some & \textbf{No} \\
Tok.\ / query   & {$\sim$}2--5K & {$\sim$}1K & {$\sim$}5K & \textbf{{$\sim$}1K} \\
Auto-sync       & Varies & N/A & Manual & \textbf{Yes} \\
License         & Comm. & Apache & Mixed & \textbf{MIT} \\
\bottomrule
\end{tabularx}
\end{table}

\subsection{System Performance}
\label{sec:eval-performance}

Table~\ref{tab:performance} reports performance measurements on an Apple M3 Pro (macOS).

\begin{table}[ht]
\caption{System performance benchmarks.}
\label{tab:performance}
\small
\begin{tabularx}{\columnwidth}{@{}Xr@{}}
\toprule
\textbf{Operation} & \textbf{Time} \\
\midrule
Fresh index (49K nodes, 196K edges) & {$\sim$}6\,s \\
Fresh index (Linux kernel, 2.1M nodes, 4.9M edges) & {$\sim$}3\,min \\
Incremental re-index & {$\sim$}1.2\,s \\
Cypher query (relationship traversal) & $<$1\,ms \\
BFS call-path tracing (depth=5) & {$\sim$}0.3\,ms \\
Name search (regex) & $<$10\,ms \\
Dead code detection & {$\sim$}150\,ms \\
\bottomrule
\end{tabularx}
\end{table}

The Django repository (49,398 nodes, 196,022 edges) serves as the primary benchmark. At the upper end of scale, indexing the Linux kernel (28\,M lines of code, 75K files) produces 2.1\,M nodes and 4.9\,M edges in approximately 3 minutes, demonstrating that the pipeline scales to very large codebases. Incremental re-indexing via XXH3 content-hash comparison achieves approximately four times speedup over full re-indexing.

\subsection{Community Adoption}
\label{sec:eval-adoption}

As a proxy for practical relevance, we report early adoption metrics from the public GitHub repository. Within four weeks of the initial release (February 25, 2026), the project accumulated more than 900 stars and approximately 100 forks. Referring traffic originated primarily from Reddit (1,288 views), LinkedIn (441), and direct GitHub discovery (869). The tool is automatically detected by 10 coding agents including Claude Code, Codex CLI, Gemini CLI, Zed, and VS Code. These figures suggest that structural code retrieval addresses a concrete need among LLM agent practitioners.

\section{Discussion}
\label{sec:discussion}

\subsection{Graph vs.\ Text: When Does Structure Help?}
\label{sec:disc-when}

Our head-to-head evaluation reveals a clear division of labor. The MCP Agent excels at cross-file structural queries, hub detection, caller ranking, and dependency chain traversal, where pre-materialized graph edges avoid the linear token cost of iterative file exploration. The ten times token reduction and 2.1 times fewer tool calls translate directly into lower latency and cost. However, the Explorer Agent retains an advantage for queries requiring full source context or exhaustive pattern matching, where the graph's intentional abstraction (storing relationships but not source lines) becomes a limitation. This suggests that the optimal architecture is a hybrid: graph-based retrieval for structural queries, with fallback to file exploration for source-level tasks.

\subsection{Structural Retrieval as a Paradigm}
\label{sec:disc-paradigm}

Current RAG approaches for code primarily use embedding-based retrieval, treating code as text~\citep{zhang2023repocoder,lu2022reacc}. Our work supports the emerging paradigm of \emph{structural retrieval}, graph traversal along typed relationships, as a complement. While embedding-based retrieval excels at semantic similarity, structural retrieval excels at relational queries. The RACG survey~\citep{racgsurvey2025} identifies this as a key frontier. Recent work on AST-based chunking~\citep{zhang2025cast} similarly shows that preserving syntactic structure improves retrieval quality, supporting the hypothesis that structure-aware approaches outperform flat text retrieval for code.

\subsection{Trust and Supply-Chain Security in MCP Ecosystems}
\label{sec:disc-security}

The growing adoption of MCP servers as tool providers for autonomous agents introduces a supply-chain trust problem that, to date, has received little attention in the literature. Unlike library dependencies managed by package registries with established review processes, MCP servers are typically distributed as standalone binaries from individual repositories. Users grant these binaries broad host permissions---file-system access, process spawning, network communication---yet have limited means to verify what the binary does beyond reading its source code.

This problem is amplified in agentic settings where the LLM autonomously invokes tools without per-call user approval. A compromised MCP server could silently exfiltrate code, inject malicious changes, or establish persistent backdoors across developer environments. The attack surface extends beyond the server binary itself to vendored dependencies, build pipelines, and installer scripts.

Our approach to this challenge (Section~\ref{sec:sys-security}) combines established industry security tooling---VirusTotal (Google), Windows Defender (Microsoft), ClamAV (Cisco), CodeQL, SLSA provenance, and OpenSSF Scorecard---into an automated, zero-tolerance release pipeline. We believe this represents a necessary baseline for any MCP server that requests elevated host permissions, and advocate for the MCP ecosystem to adopt similar verification standards. The absence of such standards today means that users must either trust individual maintainers or audit each tool manually---neither of which scales as the MCP tool ecosystem grows.

\subsection{Threats to Validity}
\label{sec:disc-threats}

\emph{Internal validity.} The benchmark compares two agent configurations (MCP vs.\ Explorer) on a single LLM backend (Claude Opus 4.6); results may not generalize across models or prompting strategies. Responses were graded by the first author against manually verified reference answers on a continuous 0--1 scale. The classification thresholds (PASS ${\geq}0.80$, PARTIAL $0.40$--$0.79$, FAIL ${<}0.40$) were chosen pragmatically; alternative thresholds could shift category distributions.

\emph{External validity.} While 31 languages are benchmarked, each is represented by a single repository. Macro-heavy languages (C: 0.58 quality) remain challenging because macros are not represented in Tree-Sitter ASTs. A systematic comparison against embedding-based RAG, ctags/LSP, and other graph systems remains future work.

\emph{Construct validity.} The knowledge graph captures static structure only; runtime behavior, reflection, and dynamic dispatch are not represented. The \texttt{query\_graph} tool applies a default ceiling of 100,000 rows, which may undercount in very large codebases. All performance benchmarks were measured on a single hardware configuration (Apple M3 Pro); results may differ on other platforms.

\subsection{Future Work}
\label{sec:disc-future}

Future work priorities include a controlled empirical study comparing graph-based, text-based, and embedding-based exploration across SWE-bench~\citep{jimenez2024swebench} with ablation studies, hybrid retrieval combining structural and semantic search, and system extensions for multi-repository dependency tracking and LLM-generated graph summaries at the function and module level.

A particularly promising application domain is health informatics, where domain-specific languages (DSLs) for clinical data transformation remain difficult for LLMs to explore and generate. For example, FHIRconnect~\citep{kohler2025fhirconnect} defines a DSL for bidirectional mappings between openEHR and HL7 FHIR, a task requiring precise navigation of archetype hierarchies, profile constraints, and cross-standard type correspondences. Current LLMs struggle to generate correct mapping code without structural context about the DSL's type system and transformation rules. Extending \textsc{Codebase-Memory} with grammars for such DSLs could expose mapping relationships, archetype dependencies, and profile constraints as graph-queryable entities, enabling LLM agents to reason about data integration tasks with the same structural precision demonstrated for general-purpose languages in this work.

\section{Conclusion}
\label{sec:conclusion}

\textsc{Codebase-Memory} demonstrates that treating code structure as a first-class, queryable graph, rather than text to be searched, delivers order-of-magnitude efficiency gains with competitive accuracy. A multi-phase Tree-Sitter pipeline with parallel worker pools, 6-strategy call resolution, and Louvain community detection produce a rich knowledge graph queryable in under a millisecond, exposed to any MCP-compatible agent without infrastructure overhead. Implemented as a single statically linked C binary with zero dependencies, the system scales from small projects to the Linux kernel (2.1\,M nodes in {$\sim$}3 minutes). At ten times lower token cost and 2.1 times fewer tool calls across 31 languages, structural retrieval provides a viable foundation for LLM-based code intelligence. Beyond efficiency, the system addresses the emerging supply-chain trust challenge of MCP tool ecosystems through an automated, zero-tolerance release verification pipeline integrating industry-standard antivirus scanning, build provenance, and dependency integrity checks---a level of binary verification not commonly found in open-source developer tools. Early community adoption (Section~\ref{sec:eval-adoption}) confirms practical demand for this approach.

\section*{Data Availability Statement}

The \textsc{Codebase-Memory} source code, benchmark scripts, and evaluation data are publicly available at \url{https://github.com/DeusData/codebase-memory-mcp} under the MIT license. The version evaluated in this paper corresponds to release \texttt{v0.5.5}.


\bibliographystyle{unsrtnat}
\bibliography{references}

\end{document}